# High-order galaxy correlation functions in the APM Galaxy Survey

Enrique Gaztañaga
*Department of Physics, Astrophysics, University of Oxford, Nuclear & Astrophysics Laboratory, Keble Road, Oxford OX1 3RH*

**ABSTRACT**
We estimate J-point galaxy averaged correlation functions $\overline{\omega}_J(\theta)$ for $J = 2, ..., 9$, in a sample of the APM Galaxy Survey with more than $1.3 \times 10^6$ galaxies and a depth $\mathcal{D} \sim 400\,h^{-1}\,\mathrm{Mpc}$. The hierarchical amplitudes $s_J = \overline{\omega}_J/\overline{\omega}_2^{J-1}$ are roughly constant, up to $J = 9$, between $\theta\mathcal{D} \sim 0.5\,h^{-1}\,\mathrm{Mpc}$ and $\theta\mathcal{D} \sim 2\,h^{-1}\,\mathrm{Mpc}$ and decrease slowly for larger scales. At scales $\theta\mathcal{D} > 7\,h^{-1}\,\mathrm{Mpc}$ we find strong similarities between the statistical properties of the galaxy fluctuations $\delta_g$ and the theoretical properties of matter fluctuations $\delta_m$ evolving under the influence of gravity in an expanding universe on assumption that the initial fluctuations are small and Gaussian. This is most easily explained if at large scales there is no significant biasing between matter and galaxy fluctuations, i.e. $\delta_g \simeq \delta_m$.

The comparison of the skewness in the CfA and SSRS catalogues with comparable subsamples of the APM indicates that the volume of a "fair sample" has to be much larger that the one in the combined CfA/SSRS catalogues.

## 1 INTRODUCTION

The J-point correlation functions have proved very useful for the analysis of the large scale structure. One might also use other approaches, such as the one-point probability distribution function (PDF) (e.g. Gaztañaga & Yokoyama 1993, Sutherland, Maddox & Efstathiou 1994) but correlation functions have some clear advantages. For catalogs of angular positions the correlation analysis can be used to sort out the effects of projected clustering and estimate the intrinsic three-dimensional properties. This seems more difficult (if not impossible) with the PDF. Moreover, there are analytical results for J-point correlations both in perturbation theory and in highly non-linear regime of gravitational growth (see below). Some of these models give degenerate solutions for $J < 4$, and they can only be tested using the higher order functions, e.g. equation (41) or (42). The J-point amplitudes can also be used as an expansion to the PDF (Juszkiewicz et al. 1993) with an accuracy given by the highest order $J$ for which the amplitudes are known. Thus, there are important motivations to extend the analysis of correlation functions to the largest orders available. The problem in practice is that for $J > 3$ the estimation of $\xi_J(r_1, ..., r_J)$ from galaxy catalogs is difficult as one has to work in a multidimensional space. In this paper we simplify the analysis by considering volume and area averaged correlations which gives better signal to noise properties and yet provides enough information to test the models. With this simplification we are able to estimate up to the 9th order amplitudes of $\xi_J$ at small scales and up to the 6th order at larger scales.

Observations of clustering in galaxy catalogues seem to find a particular relation or hierarchy for the J-point galaxy correlations, the so called *Hierarchical model*, which could be a trace of the gravitational nature of the evolution of large scale matter fluctuations. Early observational evidence for this hierarchy was found in angular catalogues of optically selected galaxies (e.g., Groth & Peebles 1977; Fry & Peebles 1978; Sharp et al. 1984). Recent analyses have confirmed these results and extended them to redshift catalogues and to IRAS galaxies (e.g Szapudi, Szalay & Boschan 1992; Meiksin, Szapudi & Szalay 1992; Gaztañaga 1992; Bouchet et al. 1993). From theory, hierarchical forms have been obtained in non-linear perturbation theory with initial Gaussian fluctuations (Peebles 1980, Fry 1984b, Goroff et al. 1986, Bernardeau 1992) and for scale-free models in the highly non-linear regime of gravitational clustering (e.g. Davis & Peebles 1977, Fry 1984a, Hamilton 1988). Numerical simulations also show similar properties (e.g. Efstathiou et al. 1988, Bouchet & Hernquist 1992, Weinberg & Cole 1992, Lahav et al. 1993, Fry, Melott & Shandarin 1993, Lucchin et al. 1993). On the other hand, there is a large literature on models with non-Gaussian initial fluctuations which do not have hierarchical correlations at large scales (e.g., Moscardini, et al. 1993 and references therein). Such models can display interesting behavior of the higher order moments, and observations might be able to distinguish them from models with initially Gaussian fluctuations.

In the standard *high peaks biasing* model the observable galaxy distribution is related to the underlying matter distribution assuming that galaxies form at peaks above some global threshold in the smoothed linear density field (e.g. Bardeen et al. 1986). In the limit of high threshold and small variance, this model is well approximated by the linear bias scheme, in which the smoothed galaxy $\delta_g(x)$ and mass density fields $\delta_m(x)$ are linearly related: $\delta_g(x) = b\,\delta_m(x)$. However, since galaxy formation is a complex, non-linear process involving both gravitational and non-gravitational interactions, the relation between the mass and the galaxy distributions may be more complicated than in the high peaks biasing model. Fry & Gaztañaga (1993) have shown that any local biasing $\delta_g = f(\delta_m)$ between the galaxy and the matter fluctuations preserves the hierarchical structure at large scales but changes the amplitudes. On the other hand, Frieman & Gaztañaga (1994) have shown how one could use the observed 3-point amplitude to identify the effects of non-local biasing. Thus the hierarchical amplitudes in the galaxy distribution can help us to learn about the corresponding matter amplitudes and the effects of biasing.

In this paper we present a preliminary analysis of the the J-point correlations in the APM catalogue. Section §2 contains the assumptions and the methods used in this paper while in section §3 we describe the APM sample and the results of the analysis. In section §4 we make a similar analysis over the CfA and SSRS catalogues to compare with the APM. The last section is devoted to a discussion of the implications.

## 2 METHOD OF ESTIMATION

### 2.1 Definitions and notation

A simple way to define the J-point correlation function of a density distribution is by considering the density fluctuations. We define the density fluctuation $\delta_i = \delta(\mathbf{r}_i)$ at a point $\mathbf{r}_i$ by $\rho_i \equiv \overline{\rho}\,(1 + \delta_i)$, where $\rho_i \equiv \rho(\mathbf{r}_i)$ is the density at a point $\mathbf{r}_i$ and $\overline{\rho}$ is the averaged density: $\overline{\rho} = <\rho_i>$. The statistical average, denoted by $<...>$, is over different realizations of $\delta(\mathbf{r})$ and corresponds to the average over positions in a fair sample of the universe.

Using this notation the J-point correlation functions are defined as:

$$\xi_J(\mathbf{r}_1,...,\mathbf{r}_J) \equiv <\delta_1,...,\delta_J>_c \quad (1)$$

where $<...>_c$ stands for the "connected" part of the expectation values. The "connected" part corresponds to the contribution to the probability $<\delta_1,...,\delta_J>$ which does not include any conditional probability of lower order. Up to $J=3$ we have $<...> = <...>_c$ but for $J=4$:

$$<\delta_1\delta_2\delta_3\delta_4> = <\delta_1\delta_2\delta_3\delta_4>_c + \sum_{ijkl} <\delta_i\delta_j><\delta_k\delta_l> \quad (2)$$

as the conditional probability to have a pair $i,j$ given a pair $k,l$ contributes directly to the probability $<\delta_1\delta_2\delta_3\delta_4>$.

For a point process the density $\overline{n}$ is defined by the probability, $dP_1$, that a galaxy is found to be inside a randomly placed volume element $dV$:

$$dP_1 = \overline{n}\,dV. \quad (3)$$

To relate the definitions in a continuous density field to the point process it is useful to consider the so called *Poisson model*, by which the probability $dP(\mathbf{r})$ to find a galaxy in a given volume element $dV$ around $\mathbf{r}$ is given by the local density, $\rho(\mathbf{r})$, of a continuous field:

$$dP(\mathbf{r}) = \rho(\mathbf{r})dV, \quad (4)$$

and is independent of what happened in neighboring elements. The probability to find one galaxy in a *randomly* placed element is $dP_1 = <dP(\mathbf{r})>$ and therefore, comparing equations (3) and (4), we have that $\overline{n} = \overline{\rho}$. This model gives a prescription to estimate the galaxy correlation functions. For example, the two-point correlation function, $\xi_2(r)$, is given by the joint probability $dP_2$ that two galaxies are found in the two volume elements $dV_1$ and $dV_2$ placed at separation $r = r_{12}$:

$$\begin{aligned} dP_2 &= <\rho_1\rho_2>dV_1dV_2 = \overline{\rho}^2\,[\,1+<\delta_1\delta_2>\,]\,dV_1dV_2 \\ &= \overline{n}^2\,[\,1+\xi_2(r)\,]\,dV_1dV_2. \end{aligned} \quad (5)$$

In a similar way, the three-point correlation function, $\xi_3(\mathbf{r}_1,\mathbf{r}_2,\mathbf{r}_3)$, is given by the joint probability $dP_3$ that three galaxies are found in the volume elements $dV_1$, $dV_2$ and $dV_3$:

$$\begin{aligned} dP_3 &= \overline{n}^3\,[\,1+\xi_2(12)+\xi_2(23)+\xi_2(13) \\ &\quad +\xi_3(123)\,]\,dV_1dV_2dV_3, \end{aligned} \quad (6)$$

where $\xi_2(ij) = \xi_2(r_{ij})$ and $\xi_3(123) = \xi_3(\mathbf{r}_1,\mathbf{r}_2,\mathbf{r}_3)$.

In the angular distribution the 2-point angular correlation function $w_2(\theta)$ is given in terms of the probability of finding two galaxies centered in each of the elements of solid angle $d\Omega_1$ and $d\Omega_2$ at angular separation: $\theta = \theta_{12}$:

$$d\mathcal{P}_2 = \mathcal{N}^2\,[\,1+w_2(\theta)\,]\,d\Omega_1\,d\Omega_2 \quad (7)$$

where $\mathcal{N}$ is the mean number of galaxies per unit solid angle.

For the 3-point angular correlation, $w_3(\theta,\theta_2,\theta_3)$,

$$\begin{aligned} d\mathcal{P}_3 &= \mathcal{N}^3\,[\,1+w_2(12)+w_2(23)+w_2(13) \\ &\quad +w_3(123)\,]\,d\Omega_1d\Omega_2d\Omega_3. \end{aligned} \quad (8)$$

Using correlation functions is particularly useful for angular distributions as they conveniently sort out projection effects. This can be illustrated for example in the above equation. The first term in equation (8) accounts for uncorrelated triplets that are clustered together just by chance or because of the projection. The next three terms correspond to a combination of a correlated pair with an uncorrelated galaxy that forms a triplet also by chance or projections, whereas the last term corresponds to real clustering. Thus, by studying $w_3$ instead of $d\mathcal{P}_3$ we avoid projection effects.

Consider now the (smoothed) density contrast $\delta_W(\mathbf{x})$:

$$\delta_W(\mathbf{x}) = \frac{1}{V_W}\int_S d\mathbf{x}'\,\delta(\mathbf{x}')W(\mathbf{x}-\mathbf{x}'), \quad (9)$$

$$V_W = \int_S d\mathbf{x}\,W(\mathbf{x}). \quad (10)$$

$W(\mathbf{x})$ is the window function and the integral is over all

space $\mathcal{S}$. For a top-hat window, $\delta_W(\boldsymbol{x})$ is the volume average of $\delta(\boldsymbol{x})$. In terms of the smoothed fluctuations, the volume-averaged J-point correlation functions are:

$$\overline{\xi}_J(V_W) \equiv <\delta_W^J>_c, \qquad (11)$$

Because of the fair sample hypothesis $< ... >$ corresponds to the average over positions in the sample. Therefore, $\overline{\xi}_J$ does not depend on $\boldsymbol{x}$ and it is just a function of the window and, in particular, of the volume $V_W$ in equation (10). It follows from equation (9) that:

$$\overline{\xi}_J = \frac{1}{V_W^J} \int_\mathcal{S} d\boldsymbol{r}_1...d\boldsymbol{r}_J \; W(r_1)...W(r_J) \; \xi_J(\boldsymbol{r}_1,...,\boldsymbol{r}_J). \qquad (12)$$

The case $J = 2$ is commonly used to characterize rms fluctuations $\sigma_W^2 = \overline{\xi}_2(V_W)$,

$$\overline{\xi}_2(V_W) = \frac{1}{2\pi^2} \int_0^\infty dk \; k \; P(k) \; W^2(k) \qquad (13)$$

in terms of the power spectrum of fluctuations $P(k)$.

## 2.2 Area-averaged correlations

The difficulty of estimating J-point angular correlations, $w_J(\theta_1,...,\theta_J)$, for $J > 2$, and the resulting amount of information are overwhelming as one has to consider all possible configurations in a multidimensional space. We do not necessarily need all this information. A considerable simplification can be achieved by estimating the hierarchical amplitudes from area-averaged correlations $\overline{w}_J$, which simplifies the data analysis and gives better signal to noise properties.

We define the area-averaged angular correlations $\overline{w}_J(\theta)$ in terms of the angular correlation functions $w_J(\theta_1,...,\theta_J)$:

$$\begin{aligned}\overline{w}_J(\theta) &\equiv \frac{1}{A^J} \int_A dA_1...dA_J \; w_J(\theta_1,...,\theta_J) \\ &= <\delta^J(\theta)>_c,\end{aligned} \qquad (14)$$

where $A = 2\pi(1 - \cos\theta)$ is the solid angle of the cone, $dA_J = \sin\theta_J d\theta_J d\varphi_J$ and $\delta(\theta)$ are the fluctuations inside the cone. Thus $\overline{w}_J(\theta)$ only depend on the size of the cone, $\theta$, as they correspond to smoothed correlations. In terms of the spatial correlation functions, $\xi_J(\boldsymbol{r}_1,...,\boldsymbol{r}_J)$,

$$\overline{w}_J(\theta) = \int_V d\boldsymbol{r}_1...d\boldsymbol{r}_J \; \psi(r_1)...\psi(r_J) \; \xi_J(\boldsymbol{r}_1,...,\boldsymbol{r}_J), \qquad (15)$$

where $d\boldsymbol{r}_i$ is the proper volume element (see §2.4), $\psi(r)$ is the normalized probability that a galaxy at a distance $r$ is included in the catalogue, i.e. equation (27), and $V$ is a cone of angle $\theta$ and infinite depth.

In our analysis we estimate $\overline{w}_J(\theta)$ from the central moments of the angular counts

$$m_J(\theta) \equiv \sum_{i=0}^{i=\infty} (i - \overline{N})^J \; P_i(\theta), \qquad (16)$$

where $P_i(\theta)$ is the probability of finding $i$ galaxies in a randomly selected cell of solid angle $A = 2\pi(1-\cos\theta)$ and $\overline{N} \equiv \sum_i i\, P_i$. Note that the galaxy density fluctuation in the cell is $\delta_g = (i - \overline{N})/\overline{N}$, and therefore

$$m_J = \overline{N}^J \; <\delta_g^J>. \qquad (17)$$

To obtain the correlation functions $\overline{w}_J$ we have to estimate the "connected" moments $\mu_J \equiv \overline{N}^J <\delta_g^J>_c$. The relations between $m_J$ and $\mu_J$ up to $J = 9$ are given in the Appendix A1. Because of the discreteness $<\delta_g^J>_c$ is not a good estimator of $\overline{w}_J$ unless $\overline{N} \gg 1$. A better estimator of $\overline{w}_J$ is given in the Appendix A1.

## 2.3 Hierarchical amplitudes

In the hierarchical model all high-order correlations can be expressed in terms of the two-point correlation function:

$$\xi_J(r_1,...,r_J) = \sum_\alpha Q_{J,\alpha} \sum_{ab} \prod^{J-1} \xi_2(r_{ab}). \qquad (18)$$

In graphical notation, associated with each term in equation (18) there is a graph, such that vertices, or nodes, correspond to the points $r_1,...,r_J$, and edges, or lines, between node $a$ and node $b$ correspond to factors $\xi_2(r_{ab})$ that connect all points. Thus the hierarchy is composed of "tree" graphs (connected with no cycles) of $J$ vertices and $J-1$ edges. The sum over $\alpha$ denotes topologically distinct graphs; the sum over $ab$ is over relabelings within $\alpha$. At each order $J$ there are in total $J^{J-2}$ terms, corresponding to all possible reassignments of the labels $a, b = 1, ..., J$. Sometimes the labels $\alpha$ in $Q_{J,\alpha}$ are omitted and $Q_J$ is the average over different topologies:

$$Q_J \equiv \frac{1}{J^{J-2}} \sum_\alpha N_\alpha Q_{J,\alpha}, \qquad (19)$$

where $N_\alpha$ is the number of graphs with topology $\alpha$.

For the average correlations $\overline{\xi}_J(R)$ in a spherical cell of volume $V = 4/3\pi R^3$, i.e. equation (12),

$$\overline{\xi}_J(R) = \frac{1}{V^J} \int_V d\boldsymbol{r}_1...d\boldsymbol{r}_J \; \xi_J(\boldsymbol{r}_1,...,\boldsymbol{r}_J). \qquad (20)$$

the hierarchy (18) translates into:

$$\overline{\xi}_J(R) = S_J \left[\overline{\xi}_2(R)\right]^{J-1}, \qquad (21)$$

where $S_J$ are related to $Q_J$ by:

$$S_J = B_J J^{J-2} Q_J. \qquad (22)$$

Parameters $B_J$ are estimated in Appendix A2. For the area-averaged correlations (15) we define:

$$s_J(\theta) \equiv \frac{\overline{w}_J(\theta)}{[\overline{w}_2(\theta)]^{J-1}}. \qquad (23)$$

The relations between $s_J$ and $Q_J$ or $S_J$ follow after a careful study of projection effects.

## 2.4 Projection effects & selection function

To estimate the spatial properties of density fluctuations we use *angular* positions of galaxies over a "complete" magnitude limited survey. The analysis takes into account two different projection effects. One is to distinguish between real clustering and clustering from projections. Angular correlation functions conveniently sort out this problem (see section §2.1). The other projection effect comes from the fact that catalogues are limited by apparent magnitude which introduces a selection function.

**Table 1.** Projection factors for different slopes $\gamma$ and parameters in the luminosity function: $M_0^*$ and $\alpha_0$.

| $\gamma$ | $M_0^*$ | $\alpha_0$ | $r_3$ | $r_4$ | $r_5$ | $r_6$ | $r_7$ | $r_8$ | $r_9$ |
|---|---|---|---|---|---|---|---|---|---|
| 1.7 | -19.8 | -1.0 | 1.19 | 1.52 | 2.00 | 2.71 | 3.72 | 5.17 | 7.25 |
| 1.7 | -19.3 | -1.2 | 1.21 | 1.57 | 2.12 | 2.93 | 4.13 | 5.88 | 8.44 |
| 1.7 | -20.3 | -0.8 | 1.18 | 1.48 | 1.93 | 2.56 | 3.46 | 4.73 | 6.51 |
| 1.8 | -19.8 | -1.0 | 1.20 | 1.55 | 2.08 | 2.85 | 3.98 | 5.62 | 8.00 |
| 3.0 | -19.8 | -1.0 | 1.54 | 2.85 | 5.78 | 12.4 | 27.8 | 63.9 | 150 |

The metric for an homogeneous universe, in comoving coordinates $x$, is:

$$d\tau^2 = c^2 dt^2 - a(t)^2 \left[ F^{-2}(x) dx^2 + x^2 dA^2 \right], \quad (24)$$

where $dA^2 = d\theta^2 + sin^2\theta \ d\varphi^2$. The expansion factor $a = (1+z)^{-1}$ is given in terms of the redshift $z = z(x)$ which follows Hubble's law:

$$H_0 x/c = \left[ q_0 z - q_0 + 1 + (q_0 - 1)\sqrt{1 + 2q_0 z} \right]/(1+z), \quad (25)$$

where $q_0 = \Omega_0/2$ in a matter-dominated universe. The Hubble parameter $H_0$ and all units are given in terms of $h$ so that $H_0 = 100h$ km s$^{-1}$ Mpc$^{-1}$. The proper volume element is $d\mathbf{r} = a^3 F^{-1}(x) \ x^2 dx \ dA$, where

$$F(x) = [1 - (H_0 x/c)^2 (\Omega_0 - 1)]^{1/2} \quad (26)$$

is the correction for curvature.

The selection function $\psi(x)$ is the normalized probability that a galaxy at coordinate $x$ is included in the catalogue. This probability depends on $x$ because given a galaxy with absolute magnitude $M$ its apparent magnitude $m$ is a function of $x$. We make the usual assumption that there exists a universal luminosity function $\phi(q)$, so that absolute magnitudes and positions of galaxies are uncorrelated. For a catalogue with apparent magnitudes between $m_1$ and $m_2$ the selection function is:

$$\begin{aligned}
\psi(x) &= \int_{q_1}^{q_2} dq \ \phi(q) \\
q_i(x) &= 10^{-\frac{2}{5}(M_i(x) - M^*)} \quad i = 1, 2 \\
M_i(x) &= m_i - 5\log_{10} d_L(x) - 25 \\
\phi(q) &= \phi^* q^\alpha e^{-q}.
\end{aligned} \quad (27)$$

where $d_L = x(1+z)$ is the luminosity distance in $h^{-1}$ Mpc and $\phi(q)$ is the Schechter luminosity function. We choose the amplitude $\phi^*$ to be a constant (the value of which turns out to be irrelevant in our analysis) whereas $M^* = M^*(z)$ and $\alpha = \alpha(z)$ are a function of redshift $z$ to account for k-corrections and evolution. Following Maddox *et al.* (1990a) we use:

$$M^* = M_0^* + M_1^* z \quad ; \quad \alpha = \alpha_0 + \alpha_1 z \quad (28)$$

### 2.5 Projections and hierarchical amplitudes

Because of the projection and the selection function $\psi(x)$ there is a complex relation between the angular amplitudes $s_J$ in equation (23) and the corresponding $S_J$ or $Q_J$ in three dimensions. Nevertheless, given a model for the two-point correlation function $\xi_2$, it is possible to find a numerical estimate for $s_J$. The contribution from the angular average (14) is very small whereas the one from the projection can be estimated as in Groth & Peebles (1978). For the hierarchy (18) with a power-law correlation $\xi_2(r) = (r_0/r)^\gamma$ and small angles, $\theta \ll 1$ in radians,

$$s_J = r_J \ C_J \ J^{J-2} Q_J \quad (29)$$

Parameters $C_J \sim 1$ correspond to the angular average in equation (14) and are estimated in Appendix A2, whereas $r_J$ are related to the selection function $\psi$:

$$\begin{aligned}
r_J &= \frac{I_1^{J-2} I_J}{I_2^{J-1}} \\
I_k &= \int_0^\infty F \ x^2 dx \ \psi^k \ x^{(3-\gamma)(k-1)} \ (1+z)^{(3+\epsilon-\gamma)(1-k)}
\end{aligned} \quad (30)$$

The proper clustering $\xi_2(r,t)$ has been parametrized as $F\xi_2(r)(1+z)^{-(3+\epsilon)}$ where $F = F(x)$ is the correction for curvature in equation (26). We use the selection function $\psi = \psi(x)$ given by equation (27) with the parameters by Maddox *et al.* (1990a), i.e. $\epsilon = 0$ and $M_0^* = -19.8$, $M_1^* = 1$, $\alpha_0 = -1$, $\alpha_1 = -2$ in equation (28). The resulting values of $r_J$ increase with $\gamma$ and $M_0^*$ and decrease with $\alpha_0$, but do not change much within the uncertainties in the shape of the luminosity function (see also Peebles 1980, §56). This is illustrated in Table 1 where values of $r_J$ are plotted for different parameters in the selection function. For a fixed $\gamma = 1.7$, the values of $r_J$ only change by a few percent even when $M_0^*$ and $\alpha_0$ are changed by $\Delta M_0^* = 1.0$ and $\Delta\alpha_0 = 0.4$ (the values in Table 1 correspond to the combinations that give the largest changes in $r_J$). The ranges $\Delta M_0^* = 1.0$ and $\Delta\alpha_0 = 0.4$ are much larger than the observational uncertainties in the luminosity function (e.g. Efstathiou, Ellis & Peterson 1988) and have been exaggerated to illustrate how $r_J$ are insensitive to the detailed shape of $\phi$. Small variations in $\Omega_0 = 1$, $\epsilon = 0$, $M_1^* = 1$ and $\alpha_1 = -2$ produce even smaller changes. It is important to notice that although $r_J$ are unaffected by changes in $\psi$, the overall normalization of $I_k$ can change quite a lot. While the amplitude of $\xi_2$ is uncertain by 40% for $\Delta M_0^* = 1.0$ and $\Delta\alpha_0 = 0.4$ the corresponding uncertainty in $r_3$ is only 2%. This is an excellent motivation for using the values of $S_J$ as measures of clustering.

In the analysis of the APM below, variations of $\gamma$ are only important for very large scales, $\theta > 3°$, where $\gamma$ changes from 1.8 to 3.. In this case $r_J$ suffers a considerable variation, see Table 1, and equation (29) is not a good approximation.

## 3 ANALYSIS OF THE APM DATA

### 3.1 The APM Galaxy Survey

Galaxies from the APM Galaxy Survey (Maddox *et al.* 1990a-c) are selected with apparent magnitudes, $m = b_J$, between $m_1 = 17$ and $m_2 = 20$. We use equal area projection pixel maps with a resolution of 3.52 arcminutes containing over $1.3 \times 10^6$ galaxies. With the selection function above the depth, defined as the distance where the distribution of galaxy distances peaks, is $\mathcal{D} \simeq 380 \, h^{-1}$ Mpc.

Residual contamination from merged images is $f \simeq 5\%$ for our magnitude range (Maddox *et al.* 1990a). Assuming that these mergers are uncorrelated, their net effect is to dilute the clustering. The observed fluctuations, $\delta_o$, are:

$$\delta_o = (1-f)\,\delta_g + f\delta_m \quad (31)$$

where $\delta_g$ are the galaxy fluctuations and $\delta_m$ the fluctuations from the mergers. Because the mergers are uncorrelated $<\delta_m^N> = <\delta_m^N \delta_g^M> = 0$ for all $N, M$, and therefore:

$$<\delta_o^J>_c = (1-f)^J <\delta_g^J>_c \quad (32)$$

which means that the observed values of the hierarchical amplitudes $s_J^o = <\delta_o^J>_c / <\delta_o^2>_c^{J-1}$ have to be multiplied by $(1-f)^{J-2}$ to obtain the galaxy amplitudes, i.e.

$$s_J = (1-f)^{J-2} s_J^o \quad (33)$$

These factors are approximate and therefore the error-bars will be at least as large as the corrections.

### 3.2 Counts-in-cells and error estimation

Counts-in-cells are estimated for square cells of side $l$ in a range $l = 0.03° - 20°$. The smaller scale is chosen to avoid shot-noise fluctuations. Starting from a cell where $\overline{N} \simeq 1$ we select different cell sizes so that the area of the next cell is at least twice as big as the previous one. To estimate the errors we try two prescriptions.

The first prescription is to estimate the dispersion in 4 different zones within the sample. In this case, the number of test-cells we have used to obtain the counts $P_i$ is arbitrarily large. This approach is very conservative as the errors correspond to the cosmic variance in 4 smaller surveys which will be significantly larger than the cosmic variance for the full survey.

For the second prescription, we first do a partition of the map into cells of a given size. We then group the cells (at random) in subsets and estimate the variance from the results in each of the subsets. This prescription provides a way to estimate the sampling errors without introducing the cosmic variance of the smaller maps. We have checked that the number of subsets used to estimate the errors is not very important; similar results are found using 4, 8 or 20 subsets. This is because we are in fact estimating numerically the internal variance in each quantity. This estimation can also be done using the central moments $m_J$. For example, the second moment $k_2 = \overline{\omega}_2 \overline{N}^2$ in equation (A6) can be written as:

$$k_2 = \frac{1}{M} \sum_i (N_i - \overline{N})^2 - \overline{N} \quad (34)$$

where the sum extend over the $M$ cells and $N_i$ is the number of galaxies in cell $i$. If the cells are independent, the variance of $k_2$ is:

$$Var(k_2) = <k_2^2> - <k_2>^2 = \frac{m_4 - 2m_3 + m_2 - m_2^2}{M} \quad (35)$$

where $m_J$ are the central moments in equation (16). For higher order moments, $k_J$, the corresponding expressions become more complicate and depend on higher orders $m_{J'}$, $J' = 2, ..., 2J$. In practice it is better to estimate numerically the variance by using the variance of the subsets. We have checked that for the lower moments $J < 5$ both estimations agree well.

About 3% of the pixels correspond to "small holes" caused by unmatched images (satellite trails, bright stars or very bright galaxies) which are more or less randomly distributed. To estimate fluctuations we accept a test-cell provided the holes inside cover less than $\sim 30\%$ of the cell and compensate the deficit by 'filling' the hole with the local density. This assumes that the clustering of the missing part of the cell is identical to clustering in the rest of the cell, which is a rough approximation as different scales may have different clustering. Nevertheless, this approximation does not affect the results within the resolution of our analysis because each cell is a 100% bigger than the previous one.

### 3.3 Two-point correlations

We first estimate the two-point angular correlation function $w_2(\theta)$ in equation (7) from counts in the pixel maps using:

$$w_2(\theta) = \frac{\langle N_i \, N_j \rangle}{\langle N_i \rangle \langle N_j \rangle} - 1, \quad (36)$$

that is, given a cell $i$ with $N_i$ galaxies we find all cells $j$ separated by an angle $\theta + d\theta$ and estimate the number of correlated pairs. The result is shown in Figure 1, which also shows the average correlation functions $\overline{w}_2(\theta)$ estimated from moments of counts-in-cells. Error-bars correspond to the dispersion in $\overline{w}_2$ estimated from narrow stripes that divide the sample in 4 zones (same zones as used by Baugh & Efstathiou 1994). The values of $\overline{w}_2(\theta)$ are plotted as a function of the radius $\theta = l/\sqrt{\pi}$ of a circle with the area of cell. Figure 1 also shows as a continuous line $\overline{w}_2^c(\theta)$ estimated from a numerical integration of $w_2(\theta)$, i.e. using equation (14). The agreement between the two estimations indicates that square cells give very similar results to circular cells once the size of the cells are scaled to $\theta = l/\sqrt{\pi}$. For clarity, we have not shown the errors in the estimation of $w_2(\theta)$ which are quite big at large scales (see Maddox *et al.* 1990a). The "break" from the power-law is more gentle in $\overline{w}_2$ than in $w_2$ because the averaging in equation (14) mixes small with large scales.

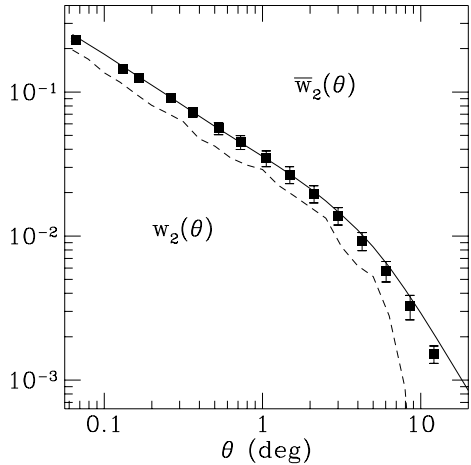

**Figure 1.** The two-point correlation $\overline{w}_2(\theta)$, squares with error-bars, estimated from counts-in-cells in the APM map compared with $w_2(\theta)$, dashed line. The continuous line correspond to $\overline{w}_2^e(\theta)$ estimated from a numerical integration of $w_2(\theta)$.

A fit of the two-point angular correlation to a power-law $w_2 \simeq A\theta^{1-\gamma}$, for scales $\theta < 2°$ gives $A \simeq 2.7 \times 10^{-2}$ and $\gamma \simeq 1.7$. Using the inversion presented in §2.5 the corresponding two-point correlation function is $\xi_2 \simeq (r_0/r)^\gamma$ with $r_0 \simeq 5\ h^{-1}$ Mpc. As mentioned above the largest uncertainty in this value is the one from the selection function.

### 3.4 Three-point averaged correlations

The three-point averaged correlation $\overline{w}_3(\theta)$ is plotted in Figure 2. Also shown is the hierarchical amplitude $s_3 = \overline{w}_3/\overline{w}_2^3$, which is divided by $10^2$ to be on the same scale. The errors are obtained from the variance in 4 different zones. The dashed line is the mean of the values in each zone, while the triangles correspond to the full map. At scales $\theta > 2°$, the mean of the zones deviates from the value in the full map which indicates that the cosmic variance in the smaller zones is important at these scales. As a consequence the errors at larger scales are too conservative.

In what follows we use our second prescription for the errors that estimates the variance in 4 subsets of the partition of the map. This prescription gives smaller errorbars at large scales. The new estimation is shown in Figure 4 below.

Figure 2 shows that $s_3(\theta)$ is roughly constant for a large range of scales; $s_3(\theta)$ only changes a factor $\sim 2$ when we change $\theta$ by a factor of $\sim 10^2$. For the smallest scales in Figure 2, i.e. $\theta < 0.1°$, the amplitude $s_3$ seems to turn over and decrease again but this effect is probably caused by shot-noise (see Appendix A1).

The three dimensional amplitude $S_3(R)$, with $R = \theta \mathcal{D}$, corresponding to $s_3(\theta)$ is given by:

$$S_3(\theta\mathcal{D}) \simeq \frac{s_3(\theta)\ (1-f)\ B_3(\gamma)}{r_3(\gamma)\ C_3(\gamma)} \qquad (37)$$

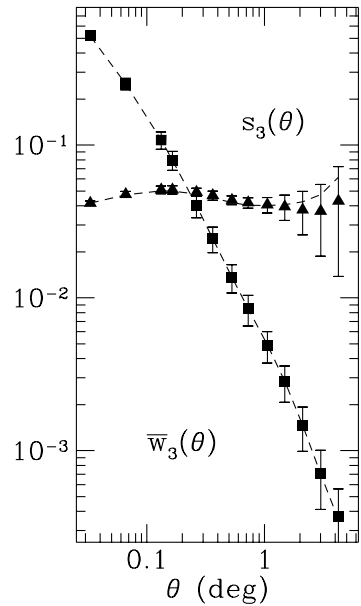

**Figure 2.** The three-point correlation $\overline{w}_3(\theta)$, squares, compared with the hierarchical amplitude $s_3(\theta) = \overline{w}_3/\overline{w}_2^2$, triangles, divided by $10^2$ to be on scale.

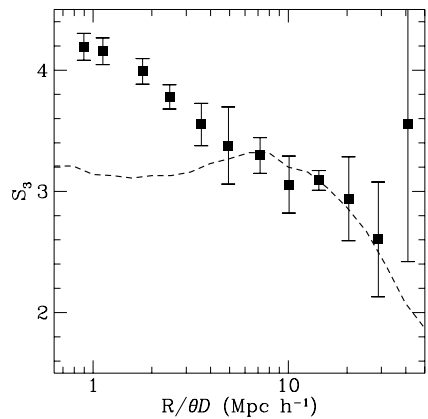

**Figure 3.** Amplitudes $S_3(R)$ for spherical cells of radius $R = \theta\mathcal{D}$, estimated from $s_3(\theta)$ in the APM. The dashed line corresponds to the amplitude $S_3$ expected for the APM in the gravitational instability scenario using second-order perturbation theory.

from equation (29) and equation (22), where $(1-f)$ is the merging correction to $s_3$, i.e equation (33). This expression is accurate for $\theta < 2°$ as the power-law model in $\xi_2$ is a good approximation (see §2.5) and the hierarchical amplitude is roughly constant. For larger scales, $\theta > 2°$, we approximate $S_3$ with equation (37) by using a local slope $\gamma = \gamma(\theta)$ (from Figure 2). The uncertainties in $s_3(\theta)$ at these large scales are larger or comparable to the error in this approximation.

*3.4.1 Comparison with perturbation theory*

In order to compare these estimations with the values in non-linear perturbation theory, we calculate $S_3(R) = \overline{\xi}_3(R)/\overline{\xi}_2(R)$ with $\overline{\xi}_2$ given by equation (13) and:

$$\overline{\xi}_3(R) = \frac{3}{(2\pi)^6} \int\int d^3k_1 d^3k_2\ B(k_1, k_2, |\mathbf{k}_1 + \mathbf{k}_2|)$$
$$W(k_1 R)\ W(k_2 R)\ W(|\mathbf{k}_1 + \mathbf{k}_2|R)\ . \qquad (38)$$

(see also Frieman & Gaztañaga 1994). For the bispectrum, $B_{123} = B(k_1, k_2, k_3)$, we use the second-order perturbation theory result:

$$B_{123} = \left[\frac{10}{7} + \left(\frac{k_1 \cdot k_2}{k_1 k_2}\right)\left(\frac{k_1}{k_2} + \frac{k_2}{k_1}\right)\right.$$
$$\left. + \frac{4}{7}\left(\frac{k_1 \cdot k_2}{k_1 k_2}\right)^2\right] P_1 P_2 + (1 \leftrightarrow 2) + (2 \leftrightarrow 3) \quad (39)$$

(Fry 1984), where the power spectrum $P_i = P(k_i)$ is the one in linear perturbation theory. We estimate $S_3$ numerically using the values of $P(k)$ obtained by Baugh & Efstathiou (1993, 1994) for the same APM sample. Of course, this measured power spectrum corresponds to the non-linear evolution, but at large scales, $R > r_0$, it provides a good approximation to the linear $P(k)$. The resulting $S_3(R)$ is plotted as a dashed line in Figure 3 where it is compared with the APM skewness. There is a good agreement for scales $5\,h^{-1}$ Mpc $< \theta \mathcal{D} < 30\,h^{-1}$ Mpc. The last point in Figure 3, at $\theta \mathcal{D} \simeq 40\,h^{-1}$ Mpc, is also consistent with the prediction after allowing for the large uncertainties in $P(k)$ at large scales, which are not plotted (see Baugh & Efstathiou 1993 & 1994). At small scales, $\theta \mathcal{D} < 5\,h^{-1}$ Mpc, the galaxy skewness is significantly bigger than the perturbation result.

The agreement in Figure 3 demonstrates that the clustering in the APM is consistent with the idea that fluctuations grow according to the gravitational instability picture. The only other assumptions that has been used in this comparison, i.e. in equation (39), is that the initial conditions were Gaussian.

### 3.5 Higher order correlations

Correlations $\overline{w}_J(\theta)$ up to $J = 9$ are shown in Figure 4 in terms of the hierarchical amplitudes $s_J = \overline{w}_J/\overline{w}_2^{J-1}$. The errors are from 4 subsets of the map. Higher order correlations $J > 9$ are dominated by poor statistics.

The hierarchical amplitudes $s_J$ in Figure 4 are roughly constant at small scales $\theta \sim 0.2°$ and decrease slowly for larger scale. For the smallest scale in Figure 4, $\theta < 0.05°$, the amplitudes seem to turn over and decrease again but this effect is probably caused by shot-noise (see Appendix).

Table 2 shows the averaged values of the amplitudes between $0.05° < \theta < 0.3°$, i.e. $0.3\,h^{-1}$ Mpc $< \theta \mathcal{D} < 2\,h^{-1}$ Mpc. The values for $s_J^{APM}$ are obtained from Figure 4 and the corresponding $S_J^{APM}$ are compensated from merging and projection effects using:

$$S_J(\theta\mathcal{D}) \simeq \frac{s_J(\theta)\,(1-f)^{J-2}\,B_J(\gamma)}{r_J(\gamma)\,C_J(\gamma)}, \qquad (40)$$

from equations (29) and (22), with $(1-f)^{J-2}$ the merging correction to $s_J$, i.e. equation (33). These relations are only valid assuming power-law correlations and constant $Q_J$. For these scales the power-law model in the two-point correlation is a very good approximation and the hierarchical amplitudes can be well approximated by constant values as shown in Figure 4. The errors in $S_J$ and $Q_J$ are the maximum between the errors from $s_J$ and the uncertainties in the merging correction.

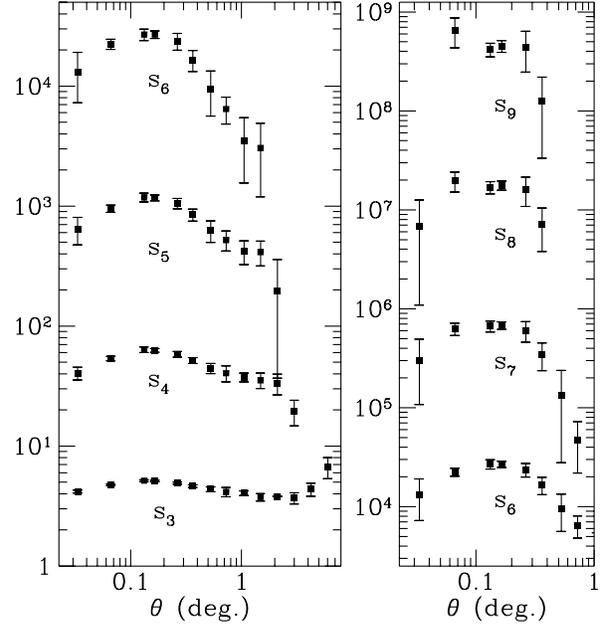

**Figure 4.** The hierarchical amplitudes $s_J = \overline{w}_J/\overline{w}_2^{J-1}$ for the APM sample. Error-bars correspond to the dispersion in 4 different zones in the APM map.

**Table 2.** The hierarchical amplitudes in the APM averaged between $0.05° < \theta < 0.3°$, i.e. $0.3\,h^{-1}$ Mpc $< \theta\mathcal{D} < 2\,h^{-1}$ Mpc.

| J | $s_J^{APM}$ | $S_J^{APM}$ | $Q_J^{APM}$ |
|---|---|---|---|
| 3 | $4.95 \pm 0.10$ | $4.04 \pm 0.08$ | $1.31 \pm 0.03$ |
| 4 | $59.2 \pm 2.4$ | $37.3 \pm 1.5$ | $2.12 \pm 0.09$ |
| 5 | $1081 \pm 59$ | $502 \pm 28$ | $3.5 \pm 0.2$ |
| 6 | $(2.5 \pm 0.1) \times 10^4$ | $(8.7 \pm 0.7) \times 10^3$ | $5.4 \pm 0.4$ |
| 7 | $(6.6 \pm 0.4) \times 10^5$ | $(1.7 \pm 0.2) \times 10^5$ | $7.4 \pm 0.7$ |
| 8 | $(1.8 \pm 0.1) \times 10^7$ | $(3.2 \pm 0.4) \times 10^6$ | $8.4 \pm 1.0$ |
| 9 | $(4.4 \pm 0.4) \times 10^8$ | $(5.8 \pm 0.8) \times 10^7$ | $7.7 \pm 1.1$ |

**Table 3.** Hierarchical amplitudes in the APM averaged at large scales, $\theta > 1°$, i.e. $\mathcal{D}\theta > 7\,h^{-1}$ Mpc.

| J | $s_J^{APM}$ | $S_J^{APM}$ | $Q_J^{APM}$ |
|---|---|---|---|
| 3 | $3.81 \pm 0.07$ | $3.16 \pm 0.14$ | $1.02 \pm 0.05$ |
| 4 | $32.5 \pm 4.2$ | $20.6 \pm 2.6$ | $1.17 \pm 0.15$ |
| 5 | $384 \pm 62$ | $180 \pm 34$ | $1.25 \pm 0.24$ |
| 6 | $3260 \pm 1340$ | $1150 \pm 580$ | $0.71 \pm 0.37$ |

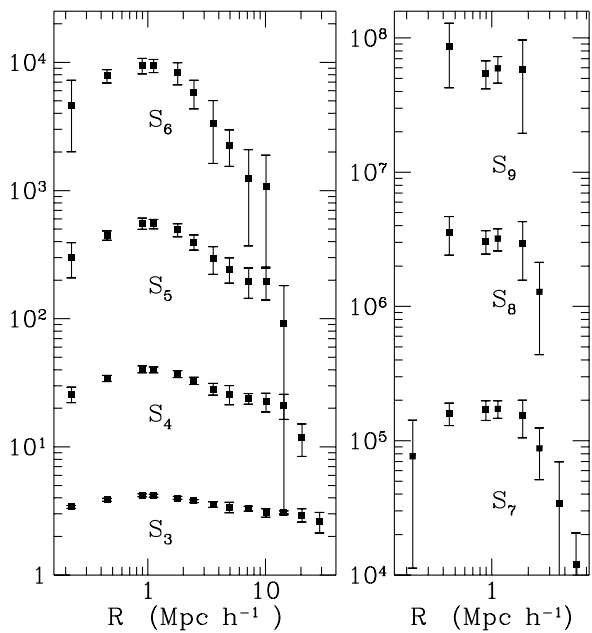

**Figure 5.** The hierarchical amplitudes $S_J = \overline{\xi}_J/\overline{\xi}_2^{J-1}$ inverted from the angular amplitudes in Figure 4.

Table 3 shows the best fit of the amplitudes in Figure 4 to a constant, weighted by the errors, for scales $\theta > 1°$, i.e. $\mathcal{D}\theta > 7\,h^{-1}$ Mpc. Notice that the number of points and the range of scales are different for each $J$. For $J = 3$ there are 8 points extending from $7\,h^{-1}$ Mpc to $80\,h^{-1}$ Mpc whereas for $J = 6$ there are only 2 points, between $7\,h^{-1}$ Mpc to $10\,h^{-1}$ Mpc. We have only use the values where the errors are not compatible with zero. The amplitudes $S_J$ and $Q_J$ are compensated for merging and projection effects using equation (40) and equation (22) with the values of $r_J$, $B_J$ and $C_J$ corresponding to the local slope $\gamma$.

Figure 5 shows the estimated values of $S_J$ from equation (40) to illustrate Tables 2-3. By approximating $Q_J$ to be constant in all the range we have probably introduced a small distortions in the shape of $S_J$ between $2\,h^{-1}$ Mpc $< \mathcal{D}\theta < 7\,h^{-1}$ Mpc, where there is a small variation of the amplitudes with scale.

### 3.5.1 Comparison with the Lick map

The results in Tables 2-3 can be compared with the averaged values of $Q_J$ from the Lick map (Shane and Wirtanen 1967). Groth & Peebles (1977) found $Q_3^{Lick} \simeq 1.25 \pm 0.18$ while Szapudi, Szalay & Boschan (1992) estimated $Q_3^{Lick} \simeq 1.44 \pm 0.07$. For $J = 4$, Fry & Peebles (1978) found $Q_4^{Lick} \simeq 3.0 \pm 0.5$ while Szapudi, Szalay & Boschan (1992) quote $Q_4^{Lick} \simeq 1.95 \pm 0.33$. These results correspond to the values averaged for all scales smaller than $\sim 5\,h^{-1}$ Mpc. But, according to our analysis, the amplitudes decrease for $\mathcal{D}\theta > 2\,h^{-1}$ Mpc and averaging could be misleading. This might explain the discrepancies between the two estimations of $Q_4^{Lick}$. For $J > 4$ the amplitudes in the Lick map found by Szapudi, Szalay & Boschan (1992) are quite uncertain, but up to $J = 6$ they seem compatible with the APM amplitudes at larger scales.

### 3.5.2 Comparison with perturbation theory

We are going to compare the amplitudes in Table 3 with the predictions in non-linear perturbation theory, as the results from N-body simulations (e.g., Efstathiou et al. 1988, Bouchet & Hernquist 1992, Weinberg & Cole 1992, Lahav et al. 1993, Fry, Melott & Shandarin 1993, Lucchin et al. 1993) prove that the perturbative approach works well even for scales where $\xi_2$ is only slightly smaller than unity, i.e. for scales larger than $r_0 \sim 5\,h^{-1}$ Mpc.

In linear perturbation theory for gravitational growth of initially Gaussian fluctuations, the dynamics are completely described by the two-point correlation $\xi_2$ or its power spectrum, $P(k)$, which grow with the scale factor. Higher order correlations are zero in linear theory as the distribution keeps Gaussian. In non-linear perturbation theory higher order correlations follow the hierarchical pattern, in equation (21), so that at each stage of the evolution $\overline{\xi}_J \simeq S_{J,m}\,\overline{\xi}_2^{J-1}$, in a self similar way. The matter amplitudes $S_{J,m}$ are characteristic of gravitational instability in an expanding universe and have been calculated analytically by Fry 1984b, Goroff et al. 1986 and Bernardeau 1992. In the quasi-linear regime $S_{J,m}$ do not vary much with time or $\Omega_0$ (Juszkiewicz, Bouchet & Colombi 1993) and therefore all dynamics are completely described by the power spectrum, just as in the linear perturbation case. Thus, $S_{J,m}$ can be used to check if fluctuations grow via gravity.

The values of $S_{J,m}$ in non-linear perturbation theory do depend slightly on the shape (but not the amplitude) of $P(k)$ and therefore on the scale. This has been illustrated in Figure 3 where $S_{3,m}$ was estimated from the measured $P(k)$. At a given scale $S_{J,m}$ can be approximated if we know the local slope $n$ in $P(k) \sim k^n$. For quasi-linear scales the observed local slope is $n \simeq -1$, which corresponds to $\gamma \simeq 2$. Thus, we can compare the amplitudes in the APM, $S_J^{APM}$ with the ones in perturbation theory $S_{J,m}$ expected for $P(k) \sim k^{-1}$. Juszkiewicz, Bouchet & Colombi (1993) found $S_{3,m} = 34/7 - (n+3)$ for a spherical top-hat window, so that $S_{3,m} \simeq 2.9$ for $n \simeq -1$, close to $S_3^{APM} \simeq 3.16 \pm 0.14$ in Table 3. A detailed comparison using the exact prediction is shown in Figure 3. Lucchin et al. (1993) have found, using numerical simulations, $S_{4,m} \simeq 16-18$ and $S_{5,m} \simeq 135-147$ for $n = -1$, which are also similar to $S_4^{APM} \simeq 20.6 \pm 2.6$ and $S_5^{APM} \simeq 180 \pm 34$ in Table 3.

Thus the amplitudes in the APM, at least up to $J = 5$, agree quite well with the predictions in the gravitational instability scenario.

## 4 COMPARISON WITH THE CFA AND SSRS

Values $S_3^{APM} \sim 3-4$ and $S_4^{APM} \sim 20-40$ in tables 2-3 are larger than the ones in the CfA and SSRS redshift catalogues: $S_3 \sim 2$ and $S_4 \sim 5$ (Gaztañaga 1992). This is probably not due to redshift distortions (Fry & Gaztañaga 1994). Is this then an intrinsic discrepancy or is it caused by the use of different techniques, e.g., the selection function?

To answer this we estimate $s_3$ for the North Zwicky Center for Astrophysics Survey (Huchra et al. 1983, hereafter CfA) and the Southern Sky Redshift Survey (da Costa et al. 1988 hereafter SSRS) using the same techniques as for the APM. In the CfA we select 1930 galaxies in a solid

## 5 DISCUSSION

The hierarchical amplitudes $s_J$ in Figure 4 are roughly constant at small scales $\theta \mathcal{D} \sim 1 \, h^{-1}$ Mpc. This result agrees with the scaling relations expected in the similarity solutions of the BBGKY model for the highly non-linear regime (e.g. Davis & Peebles 1977). The values of $Q_J$ have been estimated for the similarity solutions with different approximations. Fry (1984a) found

$$Q_J = \left(\frac{4Q_3}{J}\right)^{J-2} \frac{J}{2J-2} \tag{41}$$

whereas Hamilton (1988) estimated:

$$Q_J = \left(\frac{Q_3}{J}\right)^{J-2} \frac{J!}{2}. \tag{42}$$

Thus, to check these models we have to assume a value of $Q_3$ and compared higher order amplitudes, $J > 3$. Even after allowing for large uncertainties in the value of $Q_3$, the predictions in equations (41) or (42) are much smaller than the values of $Q_J^{APM}$ in the APM, i.e. in Table 2. One might argue that these discrepancies could be caused by small scale biasing (see below). An interesting feature shown in Table 2 is that the values of $Q_J^{APM}$ at small scales increase with the order $J$ up to $J = 9$.

Sutherland, Maddox & Efstathiou (1994) have found that at large scales the angular counts in the APM Survey fit the lognormal distribution. The lognormal distribution reproduces the Kirkwood scaling (see e.g. Coles & Jones 1991) and therefore has amplitudes $Q_J$ that increase with $\xi_2 > 1$. At large scales, $\xi_2 < 1$, the amplitudes in the lognormal distribution converge to constant values $Q_J \simeq 1$, which are similar to the APM amplitudes, $Q_J^{APM}$, in Table 3. According to perturbation theory there is nothing special about $Q_J \simeq 1$, as different values of the power index $n$ yield different values of $Q_J$. For $n \simeq -1$ we have that $Q_J \simeq 1$ is a good approximation, but at larger scales where $n > 0$, $Q_J$ are significantly lower. Thus the agreement between the lognormal distribution and perturbation theory only occurs for the small range of scales where $n \simeq -1$. In any case, the amplitudes grow much too rapidly in the lognormal distribution and at $\theta \mathcal{D} \simeq 1 \, h^{-1}$ Mpc, $Q_J$ are much larger than values we found in the APM sample.

In section §3.3 we have shown that the amplitudes $S_J^{APM}$ in Table 3, averaged at large scales, agree quite well with non-linear perturbation theory. This results agrees with the comparison of $S_3$ with second-order perturbation theory in Figure 3. There are therefore strong similarities between the properties of galaxy fluctuations and the properties of matter fluctuations evolving under the influence of gravity in an expanding universe. This indicates that the galaxy distribution has similar properties to the matter distribution: i.e. the effect of biasing on $S_{J,m}$ is small. Fry & Gaztañaga (1993) have shown that any local biasing $\delta_g = f(\delta_m) = \sum b_k \delta^k$ between the galaxy fluctuations $\delta_g$ and the matter fluctuations $\delta_m$ preserves the hierarchical structure at large scales but changes the amplitudes arbitrarily to $S_J = \mathcal{F}\left[\, b_1^{2-J} S_{J,m} \; ; \; b_2 b_1^{1-J} S_{J-1,m} \; ; \; ... \; ; \; b_{J-1} \right]$ where $\mathcal{F}$ is a known linear function of its arguments. The agreement between the matter amplitudes, $S_{J,m}$ and the galaxy amplitudes $S_J$ in the APM Survey indicates that the biasing parameters are $b = b_1 \simeq 1$ and $b_k \simeq 0$, for $k > 1$.

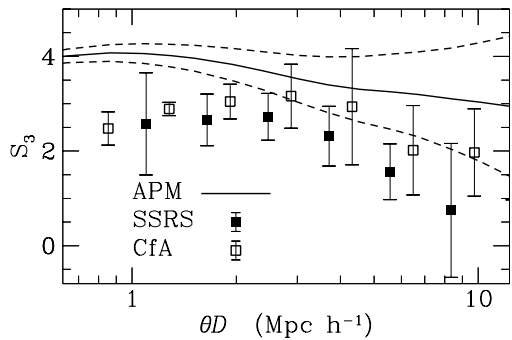

**Figure 6.** The hierarchical amplitudes $S_3$ for APM sample, continuous line, compared with the ones in the CfA sample, open squares, and the SSRS sample, closed squares. The dashed lines enclose a 70% interval of confident in the dispersion of the results of $S_3$ in 32 independent zones inside the APM, each extending over a scale of $\theta \mathcal{D} \sim 100 \, h^{-1}$ Mpc.

angle of 1.83 sr. with $m_B < 14.5$, $\delta \geq 0$ and $b \geq 40°$. In the SSRS, 1760 galaxies are selected in a solid angle of 1.75 sr. with $\log D(0) \geq 0.1$, $\delta \leq -17.5°$ and $b^{II} \leq -30°$. We use circular cells for the counts but this should not make much difference as explained above.

Results are presented in Figure 6 in terms of $S_3(\theta \mathcal{D})$ using equation (37) with $\mathcal{D}_{CfA} = 40 \, h^{-1}$ Mpc and $\mathcal{D}_{SSRS} = 50 \, h^{-1}$ Mpc, in agreement with the comparison by da Costa et al. (1988). According to the luminosity function in the CfA (Efstathiou, Ellis & Peterson 1988) the projection factor $r_3 \simeq 1.4$. Relation (37) provides a very good approximation as $\xi_2$ follows the power-law model for the scales we are interested in, i.e. $\theta \mathcal{D} < 10 \, h^{-1}$ Mpc, and the three-point amplitudes seem roughly constant. Errors in the CfA and SSRS are from 4 realizations of the cell positions. Within the errors, the CfA and SSRS results seem to agree well with the results of Gaztañaga (1992) and they have lower amplitudes than the APM.

A natural explanation for the discrepancies between the APM and the CfA/SSRS catalogue is sampling effects, i.e. that our local neighborhood is not a fair sample. After all, the APM map covers over 200 times more volume and 300 times more galaxies than the combined CfA and SSRS.

To investigate this point we divide the APM map into a 8 × 4 grid, i.e. 32 zones each extending over a scale of $\mathcal{D}\theta \sim 100 \, h^{-1}$ Mpc, about the combined depths of the CfA and SSRS. We have compared the values of $S_3$ in each independent APM zone to estimate a 70% interval of confidence in the values of $S_3$. This is displayed in Figure 6 and compared with the CfA and SSRS results, which lay within the lower end of the APM interval of confidence.

Otherwise, one would have to admit that there is a conspiracy so that after the biasing transformation the resulting galaxy amplitudes $S_J$, at each order $J$, end up being similar to the matter ones. This is a strong constraint on $b_k$ even if the agreement were only within a factor of 2, e.g. the first term in $\mathcal{F}$ for $J = 7$ constraints $b^5 < 2$ or $b < 1.15$.

To estimate $S_J$ from the angular data, $s_J$, we have assumed that absolute magnitudes and positions of galaxies are uncorrelated. Is this reasonable? In this paper we have also performed an identical analysis (to the one in the APM) using the angular positions of galaxies in the CfA and SSRS magnitude limited catalogs (§4). The results found for $S_3$ are very similar to the ones found for the same catalogs in redshift space (Gaztañaga, 1992). Moreover, they are also similar to the ones found in the angular distribution of volume limited samples (Fry & Gaztañaga 1994) which, to first approximation, are not affected by redshift distortions or luminosity-density assumptions. The agreement between all these estimations indicates that luminosity-density correlations do not affect much the values of $S_3$ or $S_4$ obtained from magnitude limited catalogs.

The values of $S_3$ inferred from the angular data in the CfA/SSRS are smaller that the ones in the APM. As argued in §2.5 it is unlikely that this discrepancy is caused by uncertainties in the luminosity function or its evolution, because the inversion coefficients, $r_J$, are quite insensitive to this. In the analysis over the Perseus-Pisces Redshift Survey by Bonometto *et al.* (1993), the values of $Q_3$ seem to decrease with increasing scales between 1 and $5\,h^{-1}$ Mpc. The corresponding variation for $S_3$ is larger than the one in Figure 3, while at large scales $S_3$ seems smaller than in the APM Survey and close to the values in the CfA/SSRS. We have shown that the small amplitudes of $S_3$ in the CfA and SSRS catalogues correspond to unlikely fluctuations in the scatter of similar APM zones (Figure 6). Thus, our analysis indicates that, to estimate high-order correlation functions, the volume of a "fair sample" has to be much larger that the one in the combined CfA/SSRS catalogues or in the Perseus-Pisces Redshift Survey. The agreement of $S_3$ in the APM (southern galactic cap) with $S_3$ in the Lick (northern galactic cap) confirms this idea and indicates that the volume of the APM is probably close to a "fair sample".

Following this sampling argument one might expect that the APM results should be closer to the results in the IRAS catalogs, which cover a bigger volume of space than the CfA or SSRS. But, in fact, Meiksin *et al.* (1992) and Bouchet *et al.* (1993) have found that $S_3$ is even smaller in the IRAS distribution. The galaxy densities in cores of clusters determined from IRAS galaxies are systematically lower than those determined from optically selected galaxies (Strauss *et al.* 1992). This effect alone can reduce the values of $S_3$ by a factor of two (Bouchet *et al.* 1993). Gaztañaga (1992) has argued that there are indications of a non-linear relation between optical and IRAS fluctuations. Thus, IRAS catalogs can not be directly compared with optical ones.

That the skewness in the nearby optical catalogues are significantly lower than in the APM Survey, might indicate that the excess of big voids found (in a "sponge-like" topology) in our local galaxy distribution, i.e. within $\mathcal{D} \sim 50\,h^{-1}$ Mpc, is not a representative feature of the Universe.

To summarise, we believe that the clustering properties found in the APM Galaxy Survey support the idea that large scale fluctuations in the universe grow according to the gravitational instability scenario and that galaxies trace matter at large scales. The values of the amplitudes found up $J = 9$ at small scales might be used to test non-linear growth and models for structure formation.

## Acknowledgements

I would like to thank Gavin Dalton and Steve Maddox for supplying the APM pixel map used in this paper. I wish to acknowledge the work by Steve Maddox, Will Sutherland, George Efstathiou and Jon Loveday in preparing the original survey. I am grateful to Carlton Baugh who provide the numerical estimation of $P(k)$ for the APM map. I would also like to thank Carlton Baugh, Gavin Dalton, George Efstathiou, Josh Frieman and Cedric Lacey for their useful comments. This work was supported by a Fellowship by the Commission of European Communities.

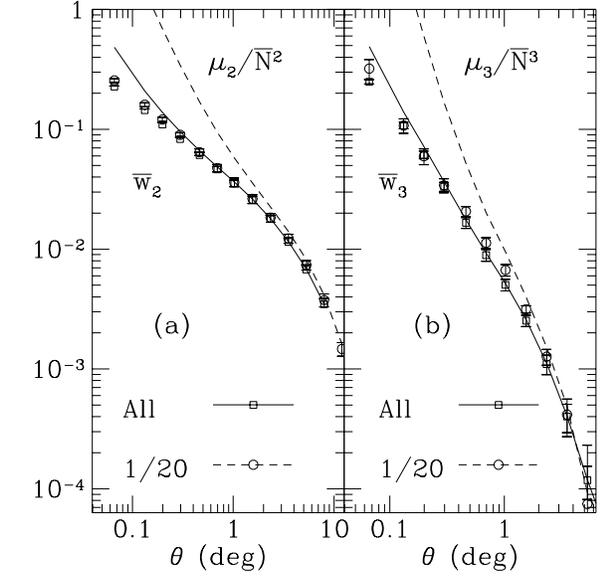

**Figure A1.** (a) Comparison of the two-point correlations with and without the discreteness correction. Squares and the continuous line correspond respectively to $\overline{w}_2$ and $\mu_2/\overline{N}^2$ in the entire map. Circles and the dashed line correspond to $\overline{w}_2(\theta)$ and $\mu_2/\overline{N}^2$ in a diluted sample. (b) Same as in (a) for the three-point correlations.

## APPENDIX A1: CONNECTED MOMENTS AND DISCRETENESS

To estimate the connected graphs, it is convenient to introduce the moment generating function $M(t)$:

$$M(t) = \sum_{J=0}^{J=\infty} \frac{m_J}{J!} t^J = <e^{t\delta}>, \tag{A1}$$

so that:

$$m_J = \left[\frac{d^J}{dt^J} M(t)\right]_{t=0} \tag{A2}$$

The connected moments are then obtained from:

$$\mu_J = \left[\frac{d^J}{dt^J} \log M(t)\right]_{t=0} \tag{A3}$$

Up to $J=9$ we find:

$$\begin{aligned}
\mu_2 &= m_2 \\
\mu_3 &= m_3 \\
\mu_4 &= m_4 - 3m_2^2 \\
\mu_5 &= m_5 - 10m_3 m_2 \\
\mu_6 &= m_6 - 15m_4 m_2 - 10m_3^2 + 30m_2^3 \\
\mu_7 &= m_7 - 21m_5 m_2 - 35m_4 m_3 + 210m_3 m_2^2 \\
\mu_8 &= m_8 - 28m_6 m_2 - 56m_5 m_3 - 35m_4^2 \\
&+ 420m_4 m_2^2 + 560m_3^2 m_2 - 630m_2^4 \\
\mu_9 &= m_9 - 36m_7 m_2 - 84m_6 m_3 - 126m_5 m_4 + 756m_5 m_2^2 \\
&+ 2520m_4 m_3 m_2 + 560m_3^3 - 7560m_3 m_2^3. \quad (A4)
\end{aligned}$$

Because of the discreteness $\mu_J = <\delta_g^J>_c \overline{N}^J$ are not good estimators of $\overline{w}_J \overline{N}^J$ unless $\overline{N} \gg 1$. A better estimator of $\overline{w}_J \overline{N}^J$ is given in terms of $k_J$ using the Poisson model introduced in §2.1. In this model, the net effect of the discreteness is that the generating function of the discrete field, $M_{Poisson}(t)$ is:

$$M_{Poisson}(t) = M(e^t - 1) \tag{A5}$$

(see Peebles 1980, §33 or Gaztañaga & Yokoyama 1993). Using this relations we find:

$$\begin{aligned}
k_2 &= \mu_2 - \overline{N} \\
k_3 &= \mu_3 - 3k_2 - \overline{N} \\
k_4 &= \mu_4 - 7k_2 - 6k_3 - \overline{N} \\
k_5 &= \mu_5 - 15k_2 - 25k_3 - 10k_4 - \overline{N} \\
k_6 &= \mu_6 - 31k_2 - 90k_3 - 65k_4 - 15k_5 - \overline{N} \\
k_7 &= \mu_7 - 63k_2 - 301k_3 - 350k_4 \\
    &\quad - 140k_5 - 21k_6 - \overline{N} \\
k_8 &= \mu_8 - 127k_2 - 966k_3 - 1701k_4 - 1050k_5 \\
    &\quad - 266k_6 - 28k_7 - \overline{N} \\
k_9 &= \mu_9 - 255k_2 - 3025k_3 - 7770k_4 - 6951k_5 \\
    &\quad - 2646k_6 - 462k_7 - 36k_8 - \overline{N} \quad (A6)
\end{aligned}$$

where terms to the right of $\mu_J$ are the shot-noise correction.

We have checked the accuracy of this shot-noise model by comparing the correlations in a given APM sample with the ones in a diluted version of the same sample, where only one out of twenty galaxies are selected at random. This is illustrated in Figure A1, where the connected moments $<\delta_g^J>_c = \mu_J/\overline{N}^J$ are quite different at small scales in the diluted sample and in the entire catalogue, whereas $\overline{\omega}_J$, obtained from $k_J$ in equations (A6), agree well for all scales. (The APM sample and the details of the analysis are described in section §3.)

### APPENDIX A2: COEFICIENTS $C_J$ AND $B_J$

Even for the case of a power-law correlation $\xi_2 \sim r^{-\gamma}$, parameters $B_J$ in equation (21) have no simple analytical expression because of the different topologies in the reassignments of labels $a,b$ in equation (18). The same happens with $C_J$ in equation (29) which are just the angular analogue of $B_J$ with $w_2 \sim \theta^{1-\gamma}$.

Two important classes of graphs can be estimated for any order. One is the "star" graphs where one of the labels in the pair is fixed, i.e. $(1,3)(2,3)...(J,3)$. The other is the "snake" graph where $(1,2)(2,3)...(J-1,J)$. For the "star" graph we find:

$$C_J^*(\gamma) = \frac{F_J}{F_2^{J-1}} \;,\; F_k = 2\int_0^1 z\,dz\, G_3^{k-1}(z) \quad (A7)$$

$$G_3(z) = \int_0^1 x\,dx \int_0^{2\pi} d\phi\,(z^2+x^2-2zx\cos\phi)^{(1-\gamma)/2}$$

which is all we need up to $J = 3$, i.e. $C_3 = C_3^*$ and $B_3 = B_3^*$. For the "snake" graphs with $J > 3$:

$$C_J^s(\gamma) = F_2^{-J+1} \int_0^1 z\,dz\, G_{J-1}(z)\, G_J(z) \quad (A8)$$

$$G_J(z) = \int_0^1 x\,dx \int_0^{2\pi} d\phi\, G_{J-1}(x) (z^2+x^2-2zx\cos\phi)^{(1-\gamma)/2}$$

The contribution for $J = 4$ is $C_4 = (12C_4^s + 4C_4^*)/16$. For $J > 4$ more graphs have to be considered in a straight forward way. In general the numerical values for $C_J$ are somewhere in-between $C_J^s$ and $C_J^*$ with the mean closer to $C_J^s$ as for a given $J$ there are more sets of labels in the "snake" configuration. The resulting $C_J$ are only slightly bigger than unity. For example for $J = 3$ we find numerically $C_3 \simeq 1.0031, 1.0087$ and $1.0166$ for $\gamma = 1.4, 1.7$ and $2.0$. In our estimations of $C_J$ we take a weighted average between the "star" and "snake" graphs. We have checked this procedure with the exact results up to $J = 5$. Because of the uncertainties in the observations a more accurate determination is not necessary. The values we find for $\gamma \simeq 1.8$ are: $C_3 \simeq 1.01$, $C_4 \simeq 1.03$, $C_5 \simeq 1.05$, $C_6 \simeq 1.07$, $C_7 \simeq 1.10$, $C_8 \simeq 1.13$ and $C_9 \simeq 1.16$.

A very similar analysis follows for $B_J$, in three dimensions. For spherical cells, $B_3$ has an analytical expression:

$$\begin{aligned}
B_3(\gamma) &= \frac{(6-\gamma)^2}{3(2-\gamma)^2}\left[\frac{46-48\gamma+17\gamma^2-2\gamma^3}{2(7-2\gamma)(9-2\gamma)}\right. \\
&\quad \left. - \frac{2^{2\gamma}\sqrt{\pi}\,\Gamma(5-\gamma)(20-13\gamma+2\gamma^2)}{512\,\Gamma(11/2-\gamma)}\right] \quad (A9)
\end{aligned}$$

Gaztañaga & Yokoyama (1993). For $J > 3$ we take a weighted average between the "star" and "snake" graphs as before. For $\gamma \simeq 1.8$ we find $B_3 \simeq 1.03$, $B_4 \simeq 1.10$, $B_5 \simeq 1.15$, $B_6 \simeq 1.25$, $B_7 \simeq 1.35$, $B_8 \simeq 1.44$ and $B_9 \simeq 1.58$.